Comment on: "The Frenkel Line: a direct experimental evidence for the new thermodynamic boundary"

V.V. Brazhkin[1] and J. E. Proctor[2]

[1] *Institute for High Pressure Physics RAS, 142190 Troitsk, Moscow, Russia*

[2] *Materials and Physics Research Group, University of Salford, Manchester M5 4WT, UK*

brazhkin@hppi.troitsk.ru, j.e.proctor@salford.ac.uk

In a recent publication (D. Bolmatov et al. *Sci.Rep.* **5**, 15850 (2015)[1]) the experimental observation of structural transformations on crossing the Frenkel line in supercritical argon is claimed. Here we show that no experimental evidence of the structural transformation was presented. The reported experimental observations which Bolmatov et al. claim as evidence of a transition across the Frenkel line are instead due to the irregularity of the experimental (*P*,*T*) path in their work.

In 2012 we have proposed a new dynamic line in the supercritical region on the phase diagram, the Frenkel line. We have shown theoretically and by computer simulation that crossing the Frenkel line on temperature increase or density (pressure) decrease results in the disappearance of shear rigidity, disappearance of positive sound dispersion, specific heat reaching $2k_B$ and the qualitative change of temperature dependence of key system properties [2-5]. However, in contrast to the first order liquid to gas phase transition up to the critical point, the density is not expected to change discontinuously upon crossing the Frenkel line. The line is universal: it separates two states at arbitrarily high pressure and temperature, and exists in systems with no critical point. The Frenkel line is unrelated to the boiling transition beneath the critical point. The location of Frenkel lines for several substances including Ar, Fe, $H_2O$, $CO_2$, $CH_4$ etc. have been subsequently calculated [6-9]. We have also showed by computer simulation that crossing the Frenkel line for the Lennard-Jones (LJ) fluid results in a rather smooth structural crossover [10].

Bolmatov et al. state [1] that they have experimentally detected sharp changes of structural parameters in fluid argon and relate the structural changes to crossing the Frenkel line. The main experimental evidence for "sharp" changes of the structural parameters (location and height of the first and second peaks of the X-ray structure factor) is presented in Figure 2b,c in [1]. The above parameters change slowly when the fluid is cooled from 480 K to 400 K but the rate of change increases drastically when the fluid is cooled from 400 K to 350 K. The authors have attributed this increase to crossing the Frenkel line. However, it is clear from the insert to Figure 1 and the data in Figure 2 in [1] that there are two discontinuities in the pressure-temperature path taken in the experiment. Indeed, pressure changes by 30% between 480 K and 400 K and by 15% between 350 K and 320 K, but it changes nearly five-fold (from 1.9 GPa to 0.4 GPa) between 400 K and 350 K. As a result, density sharply falls between 400 K and 350 K, which causes large changes of the structural parameters. Therefore the discontinuities in the temperature dependence of structural parameters co-incide with, and are the result of, the discontinuities in the pressure-temperature path taken. They are not the result of crossing the Frenkel line. The results of theoretical modelling (Figure 3c,d in [1]) reproduce structural parameters at the experimental temperature and pressure. The modelling data along the isochore in a wide temperature range (Fig. 4,5) basically repeat the previously published data [10] and demonstrate a smooth crossover in structural paramters at the Frenkel line without sharp anomalies.

If plotted on isochores (figure 4 in [1]), the data of Bolmatov et al. change smoothly in the pressure and temperature range considered and with no anomalies. This is consistent with earlier structural study of fluid Ar [11] where the authors have concluded that all structural parameters change smoothly. Just as with the experimental data, the sole reason for the observed discontinuities in the results of the modelling as presented in figure 3 of [1] is the discontinuities in the pressure-temperature path which is modelled.

Bolmatov et al. subsequently state that "Clearly, the mere *linear* variation of the *S*(*q*) peak positions as a function of pressure cannot affect the complex evolution of the peak positions presented in the Fig. 2 (B,C)". In fact, given that pressure changes 10-15 times faster in the temperature range 350-400 K as compared to other temperatures, the observed *linear* variation of peak positions of *S*(*q*) as a function of pressure explains the complex evolution of the peak position along the discontinuous (P,T) path taken very well.

Bolmatov et al. state that changing temperature across the Frenkel line can result in sharp structural changes whereas changing pressure cannot. This is incorrect: a line on the phase diagram can be crossed in any direction, and if the crossover or transition are seen along a given continuous path, they will also be detected along any other continuous path. Similarly, if a crossover or transition is not observed along a given continuous path then it will not be observed along any other continuous path.
In addition (figures 2 and 3 in [1]) Bolmatov et al. claim to have observed a discontinuity in the *S(q)* peak positions (not just their height and width) as a result of crossing the Frenkel line. A discontinuity in the *S(q)* peak positions can only occur if there is a discontinuity in the variation of the density of the sample with pressure and/or temperature. There is no known mechanism for the density of a sample not to vary smoothly as a function of temperature and pressure so far beyond the critical point. The density is not expected to change abruptly upon crossing the Frenkel line.

Not only is the data presented in [1] not evidence for the existence of the Frenkel line, but the pressure and temperature path taken in [1] is not in the region where the Frenkel line must lie. By definition, the Frenkel line corresponds to $c_v=2k_B$, where $c_v$ is the constant-volume specific heat [6]. We have taken the data of $c_v$ from the National Institute for Standards and Technology database (see http://webbook.nist.gov/chemistry/fluid) at pressure and temperature conditions used in [1]. This gives the range of $c_v$ between $2.8k_B$ and $2.25k_B$, significantly larger than $c_v=2k_B$ at the Frenkel line. Therefore, a crossover at the Frenkel line could not have been detected.

In summary, no direct experimental evidence for a new thermodynamic boundary is presented in [1].

Acknowledgments**:**

The authors wish to thank Kostya Trachenko, Dean Smith and Andrei Sapelkin for valuable discussions. VVB is grateful to RSF (14-22-00093) for the financial support.
Author contributions:
VB designed the research, VB and JP wrote the manuscript, all authors discussed and commented on the manuscript.
Competing financial interests:
The authors declare no competing financial interests.